# Landau Quantization of Massless Dirac Fermions in Topological Insulator

Peng Cheng[1,*], Canli Song[1,*], Tong Zhang[1,2,*], Yanyi Zhang[1], Yilin Wang[2], Jin-Feng Jia[1], Jing Wang[1], Yayu Wang[1], Bang-Feng Zhu[1], Xi Chen[1,¶], Xucun Ma[2,¶], Ke He[2], Lili Wang[2], Xi Dai[2], Zhong Fang[2], X.C. Xie[2], Xiao-Liang Qi[3,4], Chao-Xing Liu[5], Shou-Cheng Zhang[4] & Qi-Kun Xue[1,2]

[1] *Department of Physics, Tsinghua University, Beijing 100084, China*
[2] *Institute of Physics, Chinese Academy of Sciences, Beijing 100190, China*
[3] *Microsoft Research, Station Q, University of California, Santa Barbara, CA 93106, USA*
[4] *Department of Physics, Stanford University, Stanford CA 94305, USA*
[5] *Physikalisches Institut, Universität Würzburg, D-97074 Würzburg, Germany*

\* These authors contributed equally to this work.

¶ To whom correspondence should addressed. Email: xc@mail.tsinghua.edu.cn, xcma@aphy.iphy.ac.cn

The recent theoretical prediction and experimental realization[1–13] of topological insulators (TI) has generated intense interest in this new state of quantum matter. The surface states of a three-dimensional (3D) TI such as $Bi_2Te_3$[11], $Bi_2Se_3$[12] and $Sb_2Te_3$[13] consist of a single massless Dirac cones[8]. Crossing of the two surface state branches with opposite spins in the materials is fully protected by the time-reversal (TR) symmetry at the Dirac points, which cannot be destroyed by any TR invariant perturbation. Recent advances in thin-film growth[14-16] have permitted this unique two-dimensional electron system (2DES) to be probed by scanning tunneling microscopy (STM) and spectroscopy (STS)[16-17]. The intriguing TR symmetry protected topological states were revealed in STM experiments where the backscattering induced by non-magnetic impurities was forbidden[17-19]. Here we report the Landau quantization of the topological surface states in $Bi_2Se_3$ in magnetic field by using STM/STS. The direct observation of the discrete Landau levels (LLs) strongly supports the 2D nature of the topological states and gives direct proof of the nondegenerate structure of LLs in TI. We demonstrate the linear dispersion of the massless Dirac fermions by the square-root dependence of LLs on magnetic field.

The formation of LLs implies the high mobility of the 2DES, which has been predicted to lead to topological magneto-electric effect[7,20] of the TI.

The experiments were conducted at 4.2 K in a Unisoku ultra-high vacuum low temperature STM system equipped with molecular beam epitaxy (MBE) for film growth. The stoichiometric $Bi_2Se_3$ films were prepared on graphitized SiC(0001) substrate. Details of sample preparation have been described elsewhere[15,16]. Figure 1a shows a typical STM image of the atomically flat $Bi_2Se_3$ film with a thickness of 50 quintuple layers (QL) grown by MBE. The steps on the surface are preferentially oriented along the close-packing directions and have the height (0.95 nm) of a quintuple layer. The atomically resolved STM image (Fig. 1b) exhibits the hexagonal lattice structure of the Se-terminated (111) surface of $Bi_2Se_3$[21-23]. STM images reveal a very small density of defects (approximately 1 per 50 $nm^2$) on the surface. Most of the defects (Fig. 1c) are either clover-shaped protrusions[21-23] or triangular depressions, which can be assigned to the substitutional Bi defects at Se sites or the Se vacancies, respectively. Typically, the Dirac point (DP) of the 50 QL film is about 140 meV below the Fermi level[15], resulting in a surface carrier density of $2\sim3\times10^{12}$ $cm^{-2}$. The angle-resolved photoemission spectroscopy (ARPES) data (supplementary Fig. S1) suggest that the bottom of bulk conduction band is 185 meV above the DP. Therefore, the Fermi level of the film is well within the bulk energy gap, indicating that our MBE films are of very high quality.

In STS, the differential tunneling conductance $dI/dV$ measures the local density of states (LDOS) of electrons at energy e$V$, where –e is the charge on an electron. Figure 1d shows the $dI/dV$ spectrum on the $Bi_2Se_3$ surface at zero magnetic field. The DP of the topological states corresponds to the minimum (indicated by an arrow) of the spectrum and is about 200 meV below the Fermi level. However, the actual position of the DP should be a few tens of meV higher than that in the STS. The difference is due to the tip-induced charging and will be discussed later.

The high quality of the films ensures the observation of LLs. When a uniform magnetic field is applied perpendicular to the sample surface, the energy spectrum of the two-dimensional topological states is quantized into Landau levels. Unique to the massless Dirac fermions, the energy of the n$^{th}$ level LL$_n$ has a square-root dependence on magnetic field $B$[7, 24–26]:

$$E_n = E_D + \text{sgn}(n) v_F \sqrt{2eB\hbar |n|}, \quad n = 0, \pm 1, \pm 2, ... \quad (1)$$

where $E_D$ is the energy at DP, $v_F$ is the Fermi velocity, and $\hbar$ is Planck's constant $h$ divided by $2\pi$. Here we neglect the effect of the Zeeman splitting which only leads to a correction of several meV to the Landau level spectrum for the magnetic field range (B≤11T) we used. The number of electrons per unit area on a Landau level is proportional to the magnetic field and given by

$$n_L = Z \frac{eB}{2\pi\hbar}. \quad (2)$$

The factor $Z$ is the degeneracy of the level and depends on the materials. In the case of graphene, $Z=4$ because of the two spin states and two in-equivalent Dirac cones[24]. By contrast, $Z=1$ is expected for the topological surface states of a TI with a single Dirac cone and the unconventional spin structure. Unlike the 2DES with a parabolic dispersion, the unusual LLs of massless Dirac fermions are not equally spaced. In particular, the zero-mode level LL$_0$ has anomalous properties. Leaving the n=0 level empty or completely filling it gives rise to the quantum Hall effect (QHE) with Hall conductance -1/2 or +1/2 in units of $e^2/h$, and this effect is crucial for the topological magneto-electric effect defining the TI[7,20]. In contrast, in a system with multiple Dirac cones, the half-quantized Hall conductance due to a single Dirac cone can be cancelled by the degeneracy factor Z, such as the case for graphene with $Z=4$[27,28].

The magnetic field dependence of tunneling conductance in Fig. 2a clearly reveals the development of well-defined LLs (the series of peaks) in $Bi_2Se_3$ with increasing field[29]. The spectra demonstrate a direct measurement of the Landau quantization of the massless Dirac fermions. Under strong magnetic field, more than ten unequally spaced LLs are explicitly resolved above the DP (for example, see the black curve for a magnetic field of 11 T in Fig. 2a). Very likely, the absence of LLs below the DP results from the involvement of the bulk valence band. This argument is supported by the much broader dispersion below DP in ARPES (see supplementary Fig. S1). We identify $LL_0$ as the peaks (indicated by the vertical dotted line at -200 mV in Fig. 2a) whose positions are almost independent of $B$. These n=0 peaks are located at the DP, which slightly shifts with magnetic field as a result of the redistribution of electron density of states due to the formation of LLs.

The Dirac fermion nature of the electrons is revealed by plotting the energies of LLs versus $\sqrt{nB}$ (Fig. 2b). Notably, the $E_n$'s in the vicinity of the Fermi level fall on a straight line, as predicted by Eq. (1). However, the energies of LLs with smaller index $n$ deviate from the linear fitting. This deviation can be understood by considering the electrostatic field between the sample surface and the STM tip (supplementary Fig. S2). The density of the induced charges in the presence of a tip is given by $\varepsilon_0 V/d$, where $V$~100 mV and $d$~1 nm are the sample bias voltage and the distance between tip and sample, respectively. Typically, the induced density of electrons is in the order of $10^{12}$ cm$^{-2}$, which is comparable to the intrinsic carrier density on the surface without a tip. Thus, the effect due to the electric field in tunneling junction is not negligible.

When the bias voltage is low (close to the Fermi level), the field-effect can be linearised, which predicts that the peak positions in *dI/dV* are simply given by a linear rescaling of the Landau level energies in Eq. (1) (see supplementary information). This explains why the LL positions near the Fermi level can be well fitted by a linear function of $\sqrt{nB}$, as shown in Fig. 2b. From the slope and the intercept of the linear fitting, one can obtain the electron density per LL as

$$n_L = \frac{e}{(\nu - 1/2)h}\left(\frac{\text{intercept}}{\text{slope}}\right)^2, \quad (3)$$

where $\nu$ is the number of LLs below the Fermi energy (1/2 comes from the contribution of $LL_0$). $\nu$ depends on the applied magnetic field and can be obtained by counting the peaks in the spectra in Fig. 2a. Equation (3) offers an independent measure of $n_L$, the number of electrons per unit area on a Landau level, which is proportional to the magnetic field (Eq. (2)). The slope of the line fitting obtained from plotting $n_L$ versus $B$ gives $Z=0.99$ (Fig. 3). The experiment provides direct proof of the nondegenerate structure of the LLs, as shown schematically in the insert of Fig. 3.

At higher bias voltage, the field-effect becomes less efficient to move the Dirac cone downwards because less DP shifting is needed to accommodate the same amount of induced charges after considering the cone-shaped dispersion of the topological states. In fact, the DP shifting can be rather slow after the electric field moves the bottom of the bulk conduction band below the Fermi level. Consequently, the apparent positions of the LLs with small n tend to stay above the linear fitting as shown in Fig. 2b because of the aforementioned nonlinearity in the field-effect. In addition, the quadratic terms in Hamiltonian can also contribute to the deviation from the linear relation. More detailed analysis of the quadratic contribution, the electron-electron interaction due to the induced charges and the shape of the probe tip would be required for a complete understanding of the LLs.

The finite size (~20 nm typically) of the STM tip introduces inhomogeneity to the 2DES and broadens the LLs. Besides this field-effect broadening, several intrinsic mechanisms also contribute to the finite quasiparticle lifetime, including defect scattering, electron-phonon scattering and electron-electron interaction. The defect scattering comes into effect when the distance between impurities in the 2DES is comparable to the magnetic length $l_B = \sqrt{\hbar/eB}$. To demonstrate the suppression of Landau quantization by defect scattering, we doped the $Bi_2Se_3$ sample with Ag atoms (Fig. 4). Compared with the undoped sample, the DP shifts downwards in energy because of the electron transfer from the Ag atoms to the substrate. At low defect density (Fig. 4a), no explicit change in the tunneling spectrum has been observed. But if we further increase the doping to a density so that the distance between defects is close to $l_B$~10 nm, the magnetic length at 11 T, the LL peaks in the spectra are suppressed considerably (Fig. 4b).

Besides the point defects, the step edges on the surface can also scatter the Dirac fermions, as demonstrated by the formation of standing waves at the edges[16,17,19]. The contribution of the step edge scattering to the finite lifetime of LLs, together with the electron-phonon and electron-electron scatterings, remains an open problem owing to the intricate tip-induced broadening. A preliminary analysis on the width of LLs can be found in the supplementary information. Note that the LL at the Fermi energy has the narrowest width $\Delta E$ of about 5 meV. The mean free path of the carriers on this LL is estimated to be $l = v_F \tau = \hbar v_F / \Delta E \sim 50$ nm, giving rise to a mobility in order of $10^3$ $cm^2V^{-1}s^{-1}$. The length $l$ is comparable to the typical size of a terrace on the surface, suggesting that the mean free path of electrons on Fermi level is probably limited by the step edge scattering. Although the mobility of the 2DES in $Bi_2Se_3$ film is one order of magnitude lower than that of graphene, the observation of quantum Hall plateaus under 10 T magnetic field in this massless Dirac Fermion system is conceivable. Further efforts are required to design a feasible procedure to fabricate the multi-terminal devices based on topological insulators for transport measurements.

**Acknowledgements:** We thank Y. B. Zhang for discussions. The work is supported by NSFC and the National Basic Research Program of China. XLQ and SCZ are supported by the US Department of Energy, Office of Basic Energy Sciences, Division of Materials Sciences and Engineering, under contract DE-AC02-76SF00515.

**Figure Captions**

**Figure 1** $Bi_2Se_3$ film prepared by MBE. **a**, The STM image (200 nm × 200 nm) of the $Bi_2Se_3$(111) film. Imaging conditions: V=4.0 V, I=0.1 nA. **b**, The atomic-resolution image of the $Bi_2Se_3$(111) surface (1 mV, 0.2 nA). Each spot in the image corresponds to a Se atom. Selenium atom spacing is about 4.14 Å. **c**, The defects on the surface of $Bi_2Se_3$. The left and the right images (both taken at -0.3 V, 0.1 nA) show the substitutional Bi defect and the Se vacancy, respectively. **d**, *dI/dV* spectrum taken on the $Bi_2Se_3$(111) surface. Set point: V=0.2 V, I=0.12 nA. The DP is indicated by an arrow. The single Dirac cone band structure of $Bi_2Se_3$ is schematically shown in the insert.

**Figure 2** Landau quantization of the topological states. **a**, Differential tunneling spectra for various magnetic fields from 0 to 11 T. All STS were acquired at a set point of 0.1 V and 0.12 nA. The bias modulation was 2 mV (rms) at 913 Hz. The curves are offset vertically for clarity. **b**, LL energies showing square-root dependence on *nB*. The straight lines are the linear fitting to the data points in the vicinity of Fermi level.

**Figure 3** The degeneracy of a LL. The straight line fits the density of electrons on a LL versus magnetic field. The slope of the line yields *Z*=0.99. The insert schematically shows the eigen-states of the Hamiltonian for topological insulator under magnetic field.

**Figure 4** Suppression of LLs by defects. The upper and lower panels show the STM images and the corresponding STS spectra of LLs at two different Ag impurity densities, respectively. An individual Ag atom is imaged as a triangular spot. The imaging conditions are 0.5 V and 0.04 nA. The LLs do not change at low density of impurities (0.0002 ML coverage) **(a)**, but are greatly suppressed **(b)** as the average distance between Ag impurities becomes ~10 nm (0.0006 ML coverage).

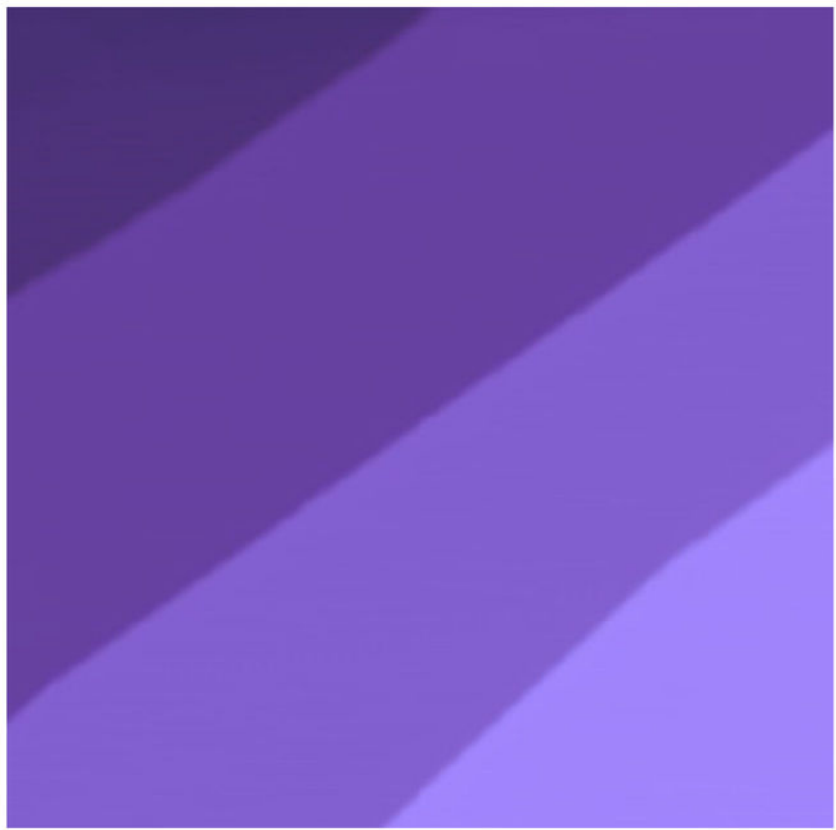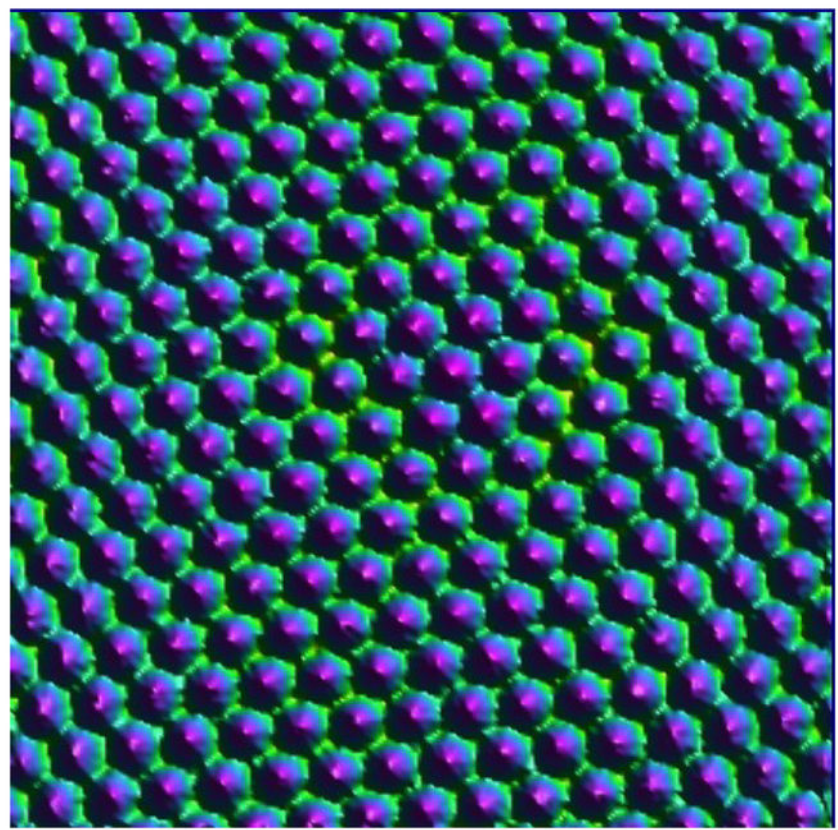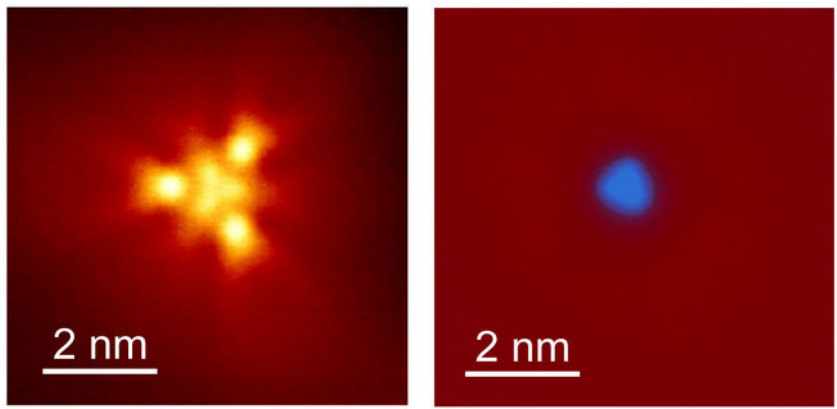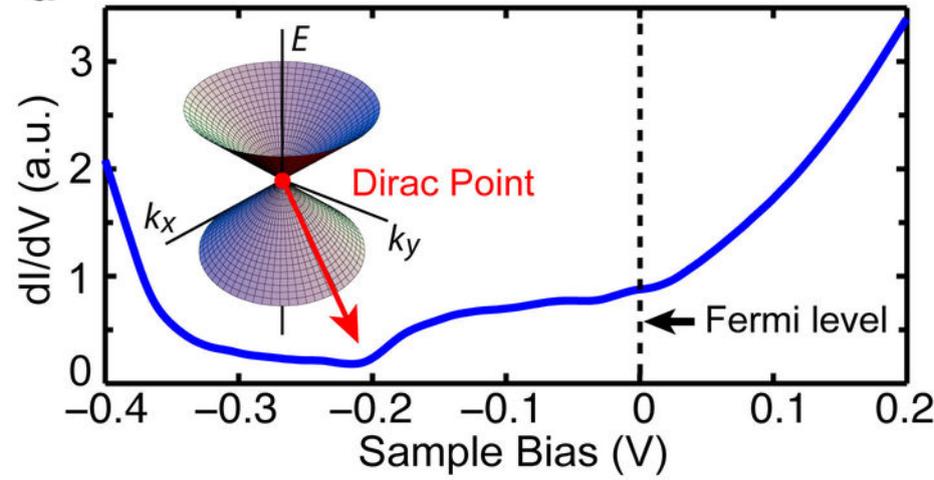

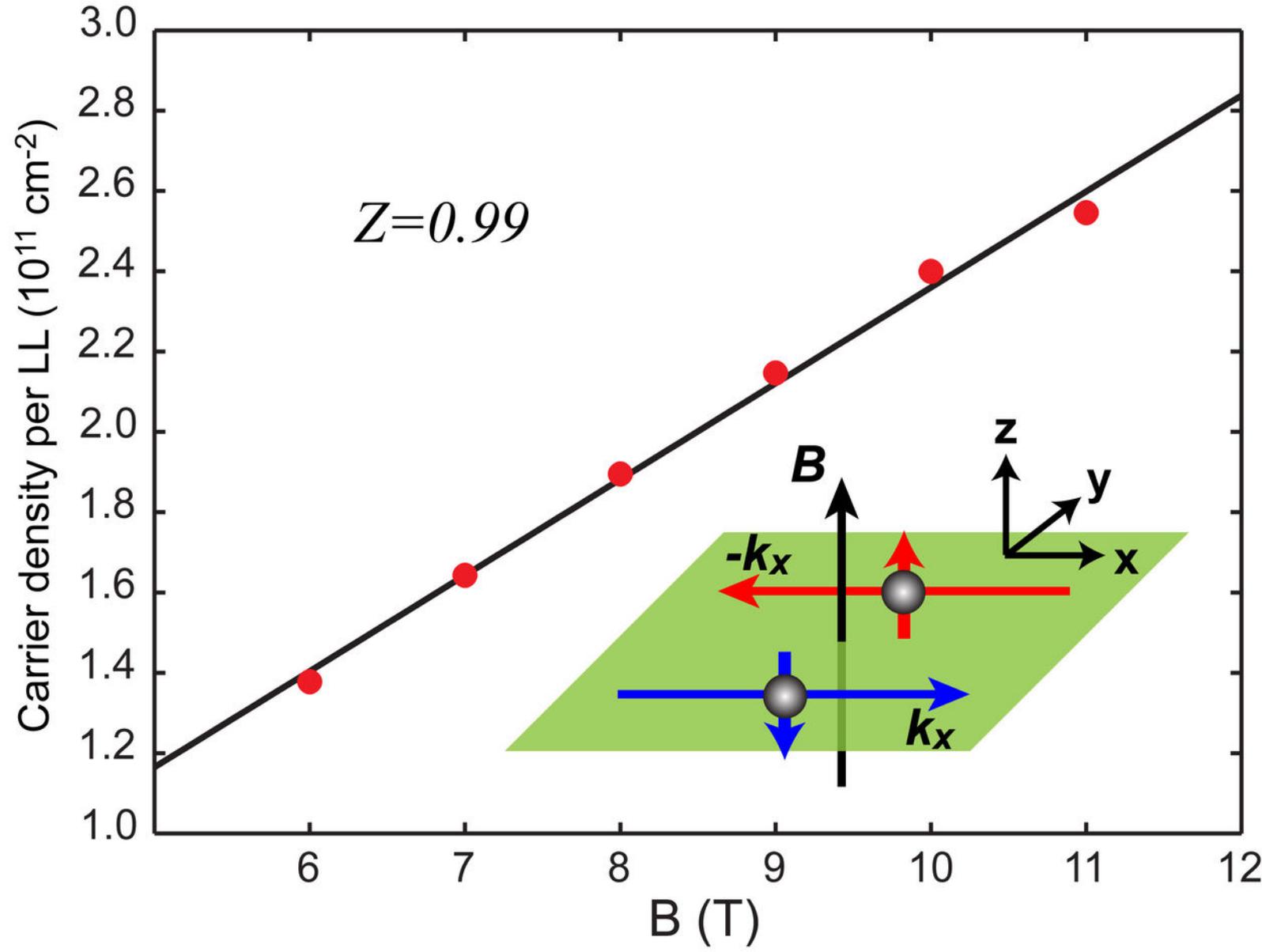

a

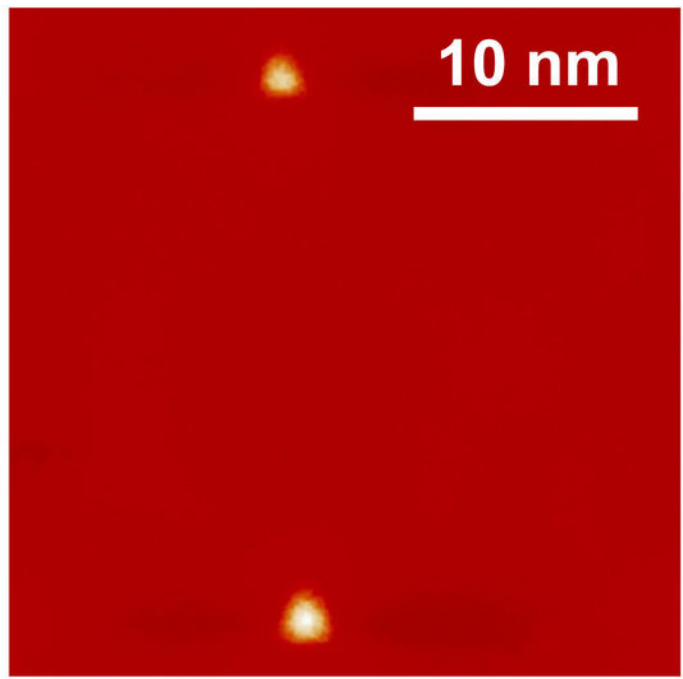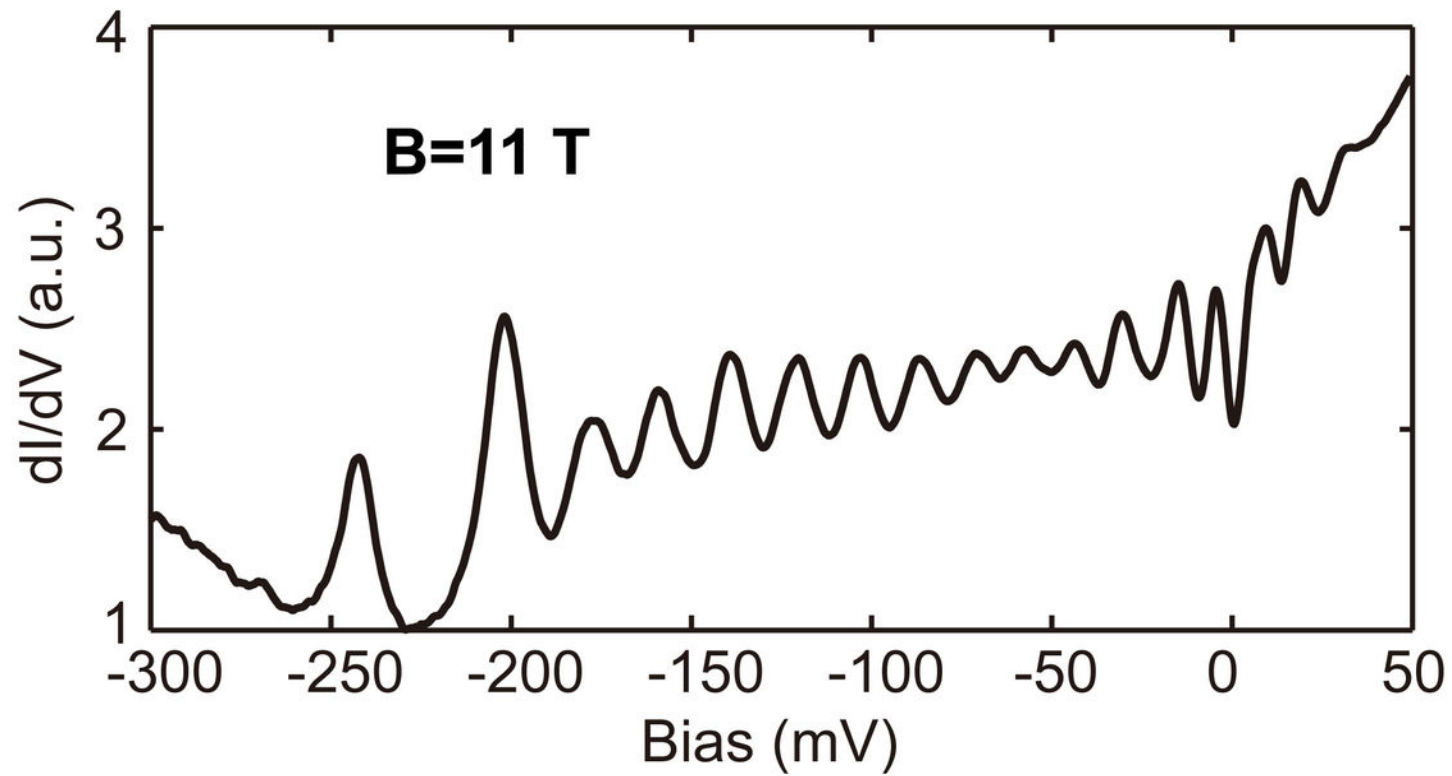

b

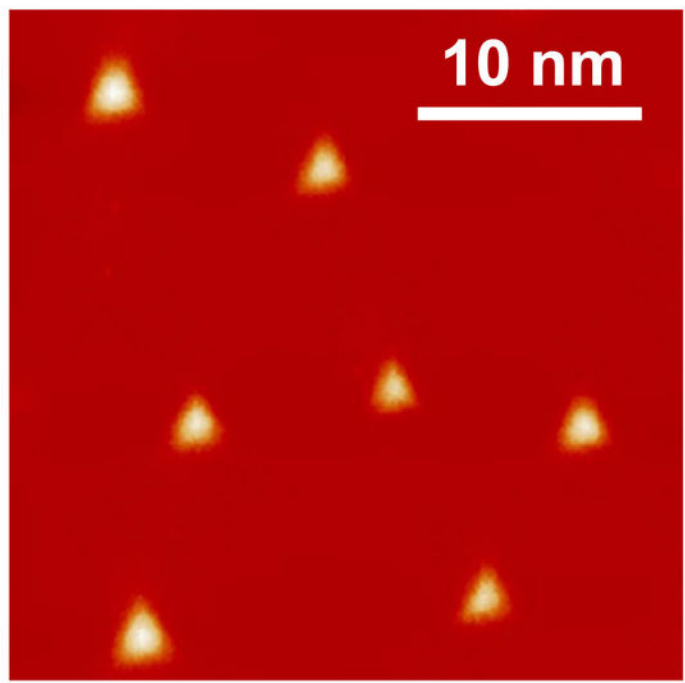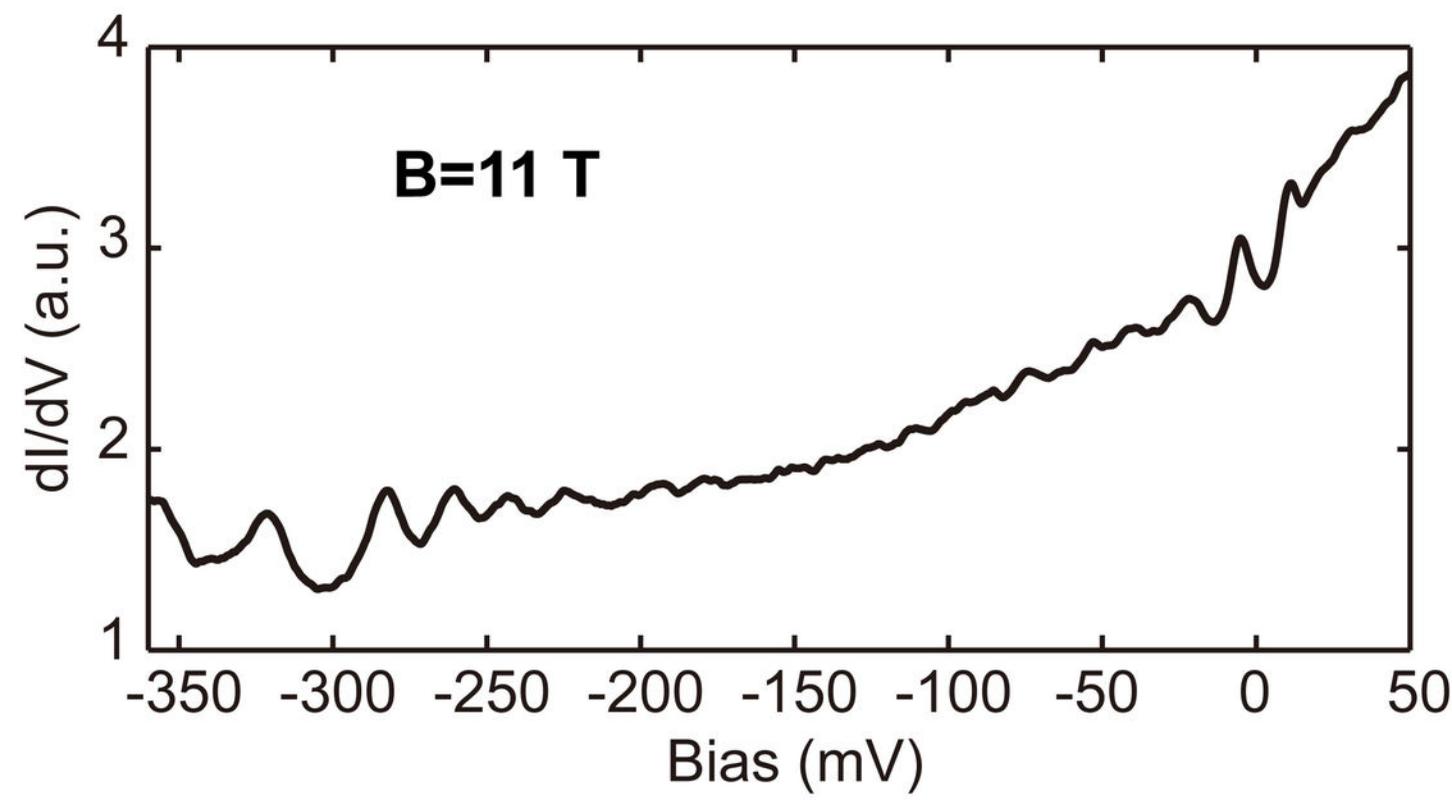

**Supplementary information for manuscript**

**"Landau Quantization of Massless Dirac Fermions in Topological Insulator"**

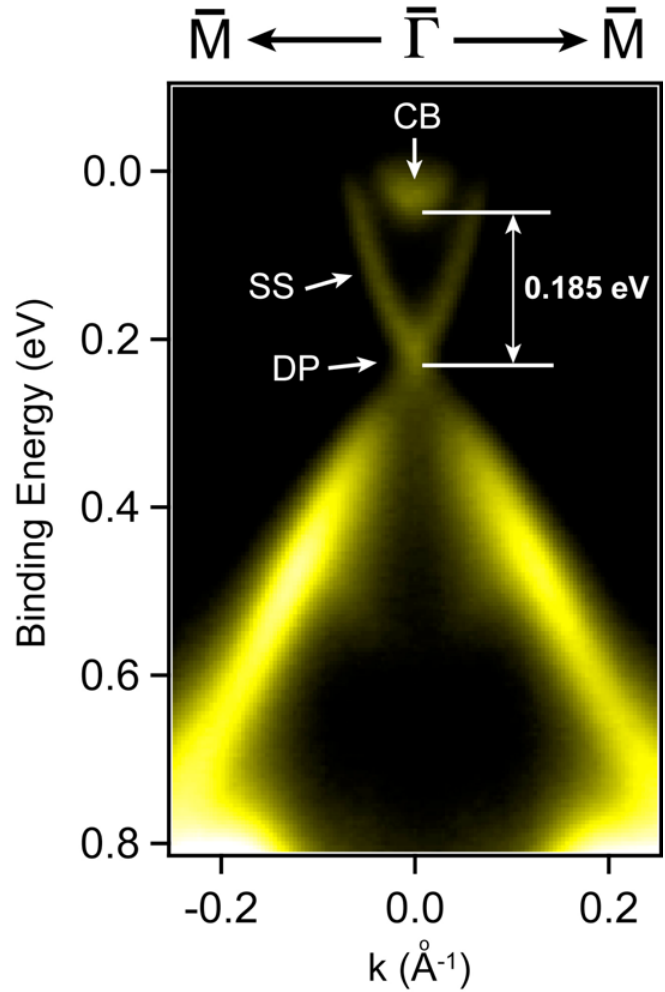

**Figure S1 | Angle-resolved photoemission spectroscopy (ARPES) of a 200 QL Bi$_2$Se$_3$ film measured at 150 K.** The film was grown on Si(111)-√3×√3-Bi surface. The growth conditions were the same as those for the film shown in Fig. 1. The photon energy used is 21.21 eV (He-Iα). The Dirac cone structure of the surface states (SS) is clearly revealed. Initially the conduction band (CB) was above the Fermi level (the zero binding energy) and invisible in ARPES. After illuminated by the 21.21 eV light at 150 K, the sample becomes electron-doped due to the surface photo-voltage (SPV) effect [see arXiv:0911.3706]. The spectra were taken after the sample had been illuminated for 6 hours when the spectra became completely stable.

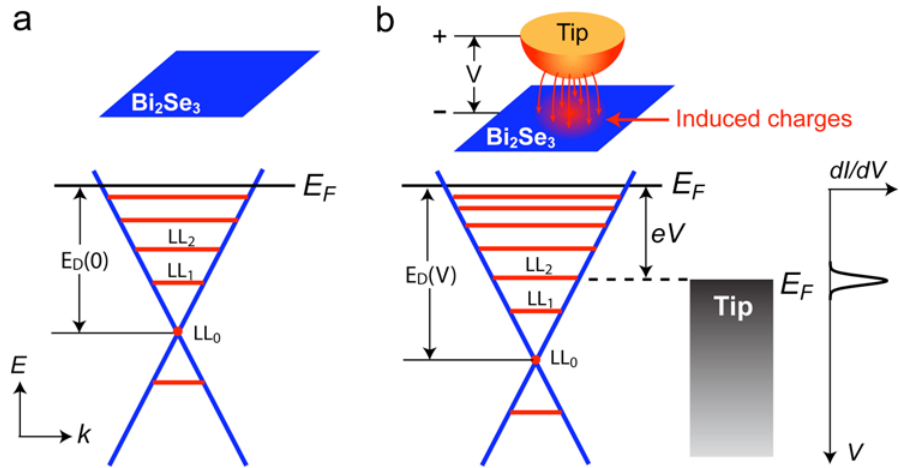

**Figure S2 | Field-effect induced by STM tip. a**, The Dirac cone without a tip. The energy of the DP is $E_D(0)$. **b**, The Dirac cone under the influence of the electrostatic field between tip and sample. For a negative bias on the sample, the DP in an area below the tip should move downwards in energy with respect to the Fermi level in order to accommodate the induced charges. Whenever the Fermi level of the tip catches up a LL during the voltage ramping, a resonance peak appears in the spectrum. Therefore, the number of peaks in the spectrum is equal to that of the LLs between $eV$ and $E_D(0)$.

## The tip-induced field effect

The number of electrons per unit area in the energy range from the DP to the Fermi level is given by

$$\sigma = \frac{E_D^2}{4\pi\hbar^2 v_F^2}. \tag{S1}$$

We refer to the Fermi level as the zero point for energy. Roughly, the surface density of induced charges is proportional to the applied bias voltage $V$ on the sample. The position of DP, as a function of bias, has to be adjusted in order to accommodate the induced charges. Therefore,

$$E_D^2(V) - E_D^2(0) \propto V. \tag{S2}$$

The Fermi level of the tip is below that of the sample when $V<0$. The above equation can be linearised when $V$ is small. Then we have

$$E_D(V) = E_D(0) + e\alpha V, \tag{S3}$$

where $\alpha$ is a constant. When the magnetic field is on, a bias voltage $V_n$ is required to see $LL_n$ and given by (see the figure below)

$$E_n - eV_n = -E_D(V_n), \tag{S4}$$

where

$$E_n = \sqrt{2n\hbar eB}\, v_F. \tag{S5}$$

From Eq. (S3)–(S5), we obtain

$$V_n = \frac{E_D(0)}{e(1-\alpha)} + \frac{\sqrt{2\hbar e}\, v_F}{e(1-\alpha)} \sqrt{nB}. \tag{S6}$$

If $V_n$ is plotted as a function of $\sqrt{nB}$, the ratio $r$ between the intercept and the slope provides an estimation of the electron density on each Landau level:

$$n_L = \frac{\sigma}{v} = \frac{e}{vh} r^2. \tag{S7}$$

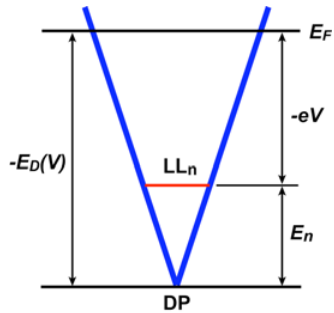

**Figure S3 | The Dirac point under the influence of tip.**

## Analysis on the peak width of LLs

There are four contributions to the width of the LLs: (1) tip-induced inhomogeneity, (2) defect scattering, (3) electron-phonon scattering, and (4) electron-electron scattering. In the present case, we suggest that the tip-induced inhomogeneity is one of the dominating factors in determining the peak width. There are two competing mechanisms of tip effect on the peak width: (1) increasing bias voltage makes the potential distribution on the surface more inhomogeneous and broaden the peaks; (2) the inhomogeneity becomes less effective at higher bias because of the reduced radius of LLs, in particular when the radius of LLs is less than that of the tip. The competition of the two effects generates a maximum in the peak width as shown in Fig. S4a. If the peak width is less than the separation between two LLs, the peaks can be easily resolved. This situation is schematically shown in Fig. S4b. In an intermediate magnetic field, for example $B_2$, there exists a region where the peak width is larger than the energy level separation. This explains the disappearance of peaks in the middle of some curves in Fig. 2a.

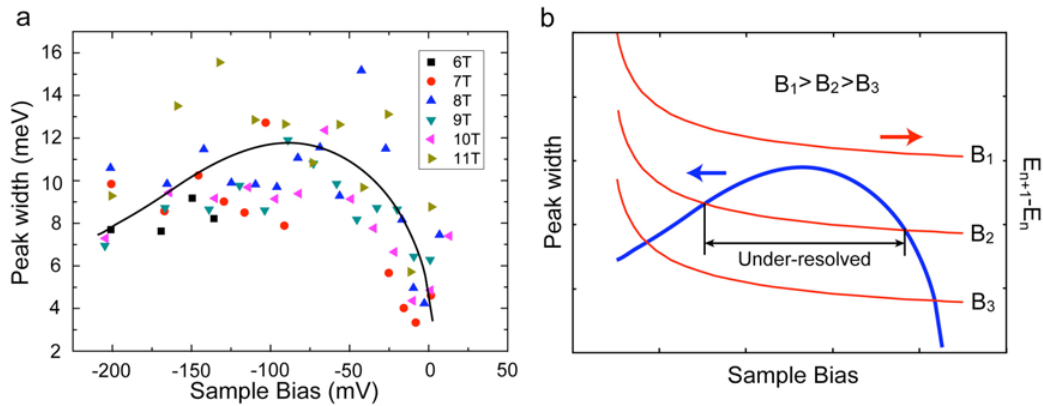

**Figure S4 | Broadening of LLs. a,** The peak width of LLs obtained by fitting the spectra with multiple Gaussians. The width reaches a maximum at about -75 mV. **b,** The resolution of spectra. At lower field, the curves for level separation always intercept with the peak width curve at two locations, leading to under-resolved regions in the *dI/dV* spectra.